\documentclass{article}
\usepackage{spconf,amsmath,graphicx}

\usepackage{cite}
\usepackage{amsmath,amssymb,amsfonts}
\usepackage{algorithmic}
\usepackage{textcomp}
\usepackage{booktabs} % To thicken table lines
\usepackage{lineno,hyperref}
\usepackage{caption}

\usepackage[dvipsnames]{xcolor}

\newcommand{\mbf}{\mathbf}

\usepackage{float}
\usepackage{gensymb}
\usepackage{textcomp}
\usepackage{etoolbox}
\usepackage{xspace}

\makeatletter
\DeclareRobustCommand\onedot{\futurelet\@let@token\@onedot}
\def\@onedot{\ifx\@let@token.\else.\null\fi\xspace}

\def\ie{\emph{i.e}\onedot}

\def\etal{\emph{et al}\onedot}
\makeatother
\title{EVALUATING DATA AUGMENTATION FOR FINANCIAL TIME SERIES CLASSIFICATION}
\name{Elizabeth Fons$^{1}$\thanks{This work was supported by the European Union's Horizon 2020 research and innovation programme under the Marie Sklodowska-Curie Grant Agreement no. 675044 (\url{http://bigdatafinance.eu/}), Training for Big Data in Financial Research and Risk Management. A. Iosifidis acknowledges funding from the Independent Research Fund Denmark project DISPA (Project Number: 9041-00004).} \qquad Paula Dawson$^{2}$ \qquad Xiao-jun Zeng$^{1}$ \qquad John Keane$^{1}$ \qquad Alexandros Iosifidis$^{3}$}

\address{$^{1}$ School of Computer Science, University of Manchester, UK.\\ % \qquad
      $^{2}$AllianceBernstein, London, UK.\\ %\qquad
      $^{3}$Department of Electrical and Computer Engineering, Aarhus University, Denmark}

\begin{document}
\ninept
\maketitle
\begin{abstract}

Data augmentation methods in combination with deep neural networks have been used extensively in computer vision on classification tasks, achieving great success; however, their use in time series classification is still at an early stage. This is even more so in the field of financial prediction, where data tends to be small, noisy and non-stationary. In this paper we  evaluate several augmentation methods applied to stocks datasets using two state-of-the-art deep learning models. 
The results show that several augmentation methods significantly improve financial performance when used in combination with a trading strategy. For a relatively small dataset ($\approx30K$ samples), augmentation methods achieve up to $400\%$ improvement in risk adjusted return performance;  for a larger stock dataset ($\approx300K$ samples), results show up to $40\%$ improvement. 

\end{abstract}

\section{Introduction}\label{sec:intro}%\vspace{-0.3cm}
Time series classification is an important and challenging problem, that has garnered much attention as time series data is found across a wide range of fields, such as weather prediction, financial markets, medical records, etc. Recently, given the success of deep learning methods in areas such as computer vision and natural language processing, deep neural networks have been increasingly used for time series classification tasks. However, unlike in the  case of image or text datasets, (annotated) time series datasets tend to be smaller in comparison, which often leads to poor performance on the classification task~\cite{Iwana2020b}. This is especially true of financial data, where a year-long of stock price data may consist of only $250$ daily prices.~\cite{Teng2020}. Therefore, in order to be able to leverage the full potential of deep learning methods for time series classification, more labeled data is needed. 

A common strategy to address this problem is  use of data augmentation techniques to generate new sequences that cover unexplored regions of input space while maintaining correct labels, thus preventing over-fitting and improving model generalization~\cite{Shorten2019}. This practice has been shown to be very effective in other areas, but it is not an established procedure for time series classification~\cite{Wen2020}~\cite{Iwana2020}. Moreover, most of the methods used are just adaptations of image-based augmentation methods that rely on simple transformations, such as scaling, rotation, adding noise, etc. While a few data augmentation methods have been specifically developed for time series~\cite{Iwana2020b, guennec2016}, their effectiveness in the classification of financial time series has not been systematically studied.

Stock classification is a challenging task due to the high volatility and noise from the influence of external factors, such as global economy and investor's behaviour~\cite{Fischer2018}. An additional challenge is that financial datasets tend to be small; ten years of daily stock prices would include   around $2500$ samples, which would be insufficient to train even a small neural network (e.g. a single-layer LSTM network with $25$ neurons has approximately $2700$ parameters). In this work we perform a systematic analysis of multiple individual data augmentation methods on stock classification. To compare the different data augmentation methods, we evaluate them using two state-of-the-art neural network models that have been used for financial tasks. As the usual purpose of stock classification tasks is to build portfolios, we compare the results of each method and each architecture by building simple rule-based portfolios and calculating the main financial metrics to assess performance of each portfolio. Finally, we analyse the combination of multiple data augmentation methods, by focusing on the best performing ones.

The contributions of the paper are as follows:
\begin{itemize}
    \item We provide the first, to the best of our knowledge, thorough evaluation of popular data augmentation methods for time series on the stock classification problem; we perform an in-depth analysis of a number of methods on two state-of-the-art neural network architectures using daily stock returns datasets. 
    \item We evaluate performance using traditional classification metrics. In addition, we build portfolios using a simple rule-based strategy and evaluate performance based on financial metrics.
\end{itemize}

The remainder of the paper is organized as follows: Section \ref{sec:relatedWork} overviews previous work on data augmentation; Section \ref{sec:da} describes the data augmentation methods used in our evaluations; Section \ref{sec:methodology} describes the experimental setup; Section \ref{sec:results} provides the experimental results; conclusions and future work are presented in Section \ref{sec:conclusions}.

\section{Related work}\label{sec:relatedWork}%\vspace{-0.3cm}
Data augmentation has proven to be an effective approach to reduce over-fitting and improve generalization in neural networks~\cite{Cui2015}. While there are several methods to reduce over-fitting in neural networks, such as regularization, dropout and transfer learning, data augmentation tackles the issue from the root, i.e., by enriching the information related to the class distributions in the training dataset. Therefore, by assuming that more information can be extracted from the dataset through augmentations, it further has the advantage that it is a model-independent solution~\cite{Shorten2019}.

In tasks such as image recognition, data augmentation is a common practice, and may be seen as a way of pre-processing the training set only~\cite{Goodfellow2016}. For instance Krizhevsky~\etal~\cite{AlexK2012} used random cropping, flipping and changing image intensity in AlexNet, Simonyan~\etal used scale jittering and flipping~\cite{Simonyan2015} on the VGG network. However, such augmentation strategies are not easily extended to time-series data in general, due to the non i.i.d. property of the measurements forming each time-series. Data augmentation has been applied to domain-specific time series data encoding information of natural phenomena with great success. Cui~\etal~\cite{Cui2015} use stochastic feature mapping as a label preserving transformation for automatic speech recognition. Um~\etal~\cite{Um2017} test a series of transformation-based methods (many inspired directly by computer vision) on sensor data for Parkinson's disease and show that rotations, permutations and time warping of the data, as well as combinations of those methods, improve test accuracy.

To date, little work has been done on studying the effect of data augmentation methods for financial data or developing methods specialized on financial time-series. For regression tasks, Teng~\etal~\cite{Teng2020} use a time-sensitive data augmentation method for stock trend prediction, where data is augmented by corrupting high-frequency patterns of original stock price data as well as preserving low-frequency ones in the frame of wavelet transformation. For stock market index forecasting, Yujin~\etal~\cite{Baek2018} propose ModAugNet, a framework consisting of two modules: an over-fitting prevention LSTM module and a prediction LSTM module.

\section{Time Series Augmentation}\label{sec:da}%\vspace{-0.3cm}
\begin{figure}
    \centering
    \includegraphics[width=\linewidth]{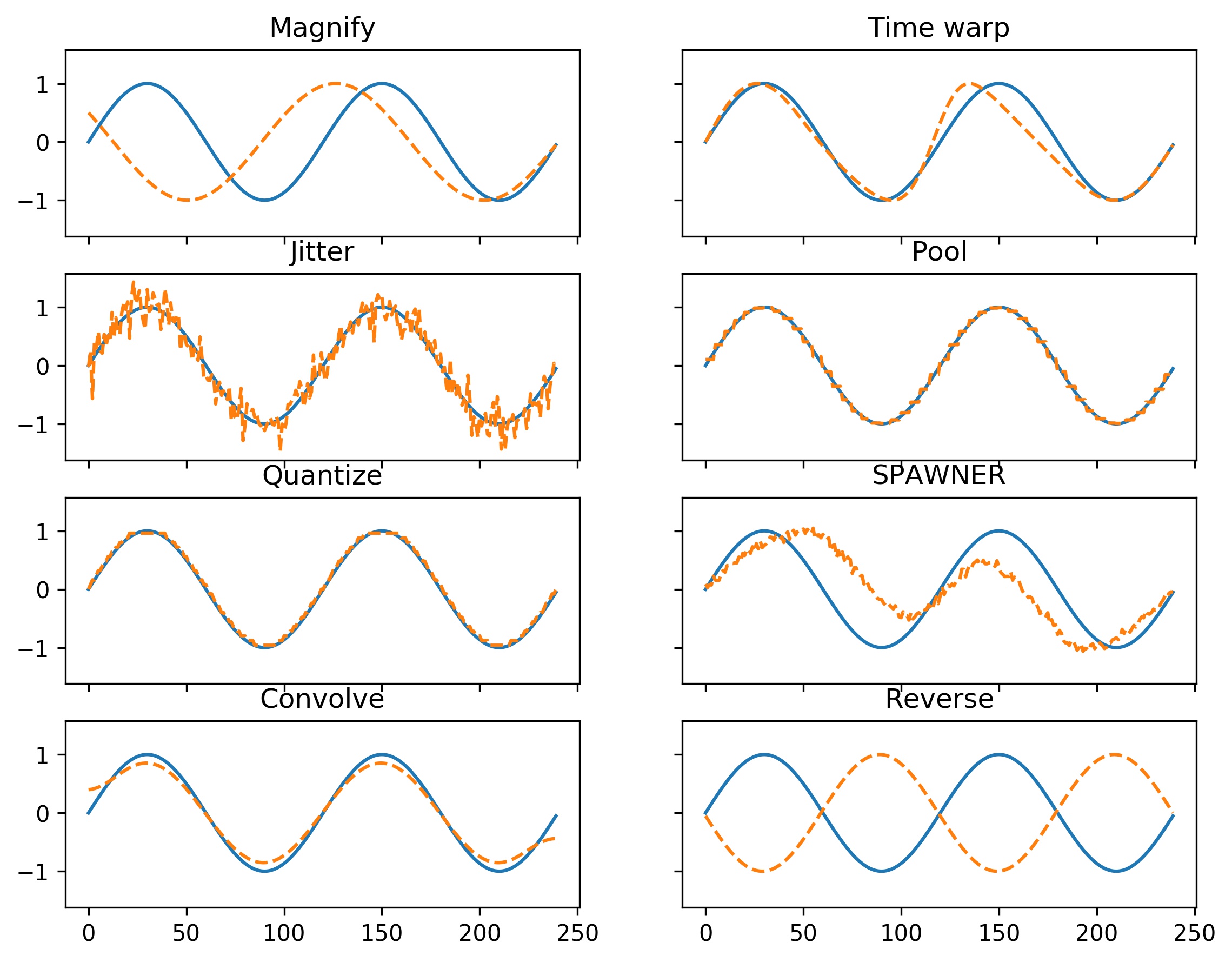}
    \caption{Examples of time-series data augmentation methods on a sine wave. The blue line corresponds to the original time-series and the dotted orange lines correspond to the generated time-series patterns.}
    \label{fig:da_ex}
\end{figure}
Most cases of time series data augmentation correspond to random transformations in the magnitude and time domain, such as jittering (adding noise), slicing, permutation (rearranging slices) and magnitude warping (smooth element-wise magnitude change). In our analysis, the following methods were used for evaluation, and examples of these transformations are shown in Figure \ref{fig:da_ex}:\\
{\bf Magnify:} a variation of window slicing proposed by Le Guennec et al~\cite{guennec2016}. In window slicing, a window of $90\%$ of the original time series is selected at random. Instead, we randomly slice windows between $40\%$ and $80\%$ of the original time series, but always from the fixed end of the time series (\ie we slice the beginning of the time series by a random factor). Randomly selecting the starting point of the slicing would make sense in  an anomaly detection framework, but not on a trend prediction as is our case. We interpolate the resulting time series to the original size in order to make it comparable to the other augmentation methods. \\
{\bf Reverse:} the time series is reversed; hence a time-series of the form $\{t_1, t_2, \ldots, t_{n-1}, t_n\}$ is transformed to $\{t_n, t_{n-1}, \ldots, t_2, t_1\}$. This method is inspired by the flipping data augmentation process followed in computer vision. \\
{\bf Jittering:} Gaussian noise with a mean $\mu = 0$ and standard deviation $\sigma = 0.01$ is added to the time series \cite{Um2017}.\\
{\bf Pool:} Reduces the temporal resolution without changing the length of the time series by averaging a pooling window. We use a window of size $3$. This method is inspired by the resizing data augmentation process followed in computer vision.\\
{\bf Quantise:} the time series is quantised to a level set $n$, therefore the difference between the maximum and minimum values of the time series is divided into levels, and the values in the time series are rounded to the nearest level \cite{quantize2000}. We used $n=25$.\\
{\bf Convolve:} the time series is convolved with a kernel window. The size of the kernel is $7$ and the type of window is Hann.\\
{\bf Time Warping:} the time intervals between samples are distorted based on a random smooth warping curve by cubic spline with four knots at random magnitudes \cite{Um2017}.\\
{\bf Sub-optimal warped time series generator (SPAWNER):} SPAWNER~\cite{spawner} creates a time series by averaging two random sub-optimally aligned patterns that belong to the same class. Following Iwana~\etal~\cite{Iwana2020b}, noise is added to the average with $\sigma=0.5$ in order to avoid cases where there is little change.

For the methods Pool, Quantise, Convolve and Time warping we used the code from Arundo \cite{arundo}\footnote{https://arundo-tsaug.readthedocs-hosted.com/en/stable/}.

%\vspace{-0.3cm}
\section{Methodology}\label{sec:methodology}%\vspace{-0.4cm}
\subsection{Datasets}%\vspace{-0.2cm}
\begin{figure}
    \centering
    \includegraphics[width=0.9\linewidth]{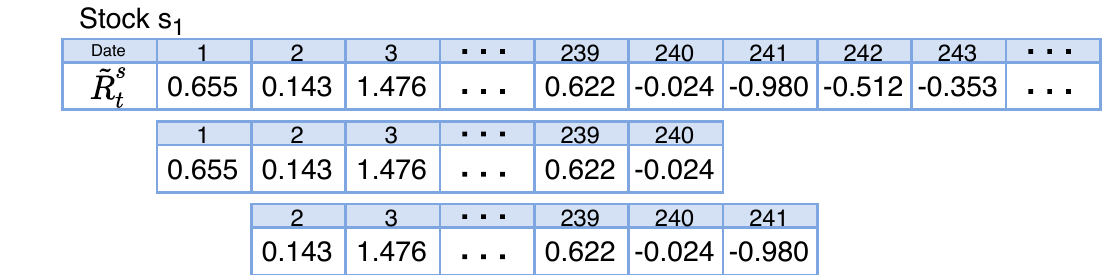}
    \caption{Construction of input sequences, segmented in $240$ time steps, with a moving window of one day.}
    \label{fig:data_diagram}
\end{figure}

\textit{Full S$\&$P500 dataset:}
The data used in this study consists of the daily returns of all constituent stocks of the S$\&$P500 index, from $1990$ to $2018$. It comprises $7000$ trading days, and approximately $500$ stocks per day. We use the data pre-processing scheme from Krauss~\etal~\cite{Krauss2017}, where the data is divided into splits of $1000$ days, with a sliding window of $250$ days. Each split overlaps with the previous one by $750$ points, and a model is trained in each one, resulting in $25$ splits in total. 
Inside each of the 25 splits, the data is segmented into sequences consisting on $240$ time steps $\{\tilde{R}^{s}_{t-239}, \ldots, \tilde{R}^{s}_{t}\}$ for each stock $s$, with a sliding window of one day, as shown in Figure \ref{fig:data_diagram}. The first $750$ days make up the training set, with the test set consisting of the last $250$ days. The training set has approximately 255K samples ((750-240)*500) and the test set has approximately 125K samples. 
The data is standardised by subtracting the mean of the training set ($\mu_{train}$) and dividing by the standard deviation ($\sigma_{train}$), i.e., $\tilde{R}^{s}_t = \frac{R^{s}_t - \mu_{train}}{\sigma_{train}}$, 
with $R^s_t$ the return of stock $s$ at time $t$. We define the problem as a binary classification task, where the target variable $Y^s_{t+1}$ for stock $s$ and date $t$ can take to values, 1 if the returns are above the daily median (trend up) and 0 if returns are below the daily median. This leads to a balanced dataset.

\noindent \textit{50 stocks dataset:} In order to have a smaller dataset, we use the same pre-processing scheme but only for the largest $50$ stocks on the S$\&$P500 measured by market capitalization, on each data split. This leads to $25500$ samples for training and $12500$ for testing.

\subsection{Augmentation}%\vspace{-0.2cm}
The training data (750 days) is divided into training and validation with a proportion $80/20$. Before splitting the data, all samples are shuffled in order to make sure that all stocks and time steps are randomly assigned to train or validation. Each train set is augmented with 1X the original size.

\subsection{Network architectures and training}%\vspace{-0.2cm}
We used two neural network architectures proposed in previous financial studies, optimizing the cross entropy loss:\\
\underline{\textit{LSTM}}: Following Krauss~\etal~\cite{Krauss2017}, we train a single layer LSTM network with $25$ neurons, and a fully connected two-neuron output. We use a learning rate of $0.001$, batch size $128$ and early stopping with patience $10$ with RMSProp as optimizer.

\noindent \underline{\textit{Temporal Logistic Neural Bag-of-Features (TLo-NBoF)}}: we adapt the network architecture proposed by Passalis~\etal~\cite{passalis2020} to forecast limit order book data. The original network was used on data samples of 15 time steps and 144 features so we adapt it for our univariate data of 240 time steps. It comprises an 1D-convolution ($25$ filters, kernel size $81$), a TLo-NBoF layer ($N_K=12$, $N_T=3$), a fully-connected layer ($50$ neurons) and a fully-connected output layer of $2$ neurons. The initial learning rate is set to $0.0005$, the learning rate is decreased on plateau of the validation loss, batch size is $256$ and the optimizer is Adam. 

\subsection{Rule-based portfolio strategy and evaluation}%\vspace{-0.2cm}
In order to evaluate if data augmentation provides an improvement in asset allocation, we propose a simple trading strategy, following the conclusions of Krauss et al~\cite{Krauss2017}. The trading rule on the full S$\&$P500 dataset is as follows: stocks in both classes are ranked daily by their predicted probability of belonging to that class, we then take the top $10$ and bottom $10$ stocks and build a long-short portfolio by equally weighting the stocks. Portfolios are analysed after transaction costs of 5 bps per trade.  

On the 50-stocks dataset, building a long-short portfolio would not be profitable as it consists of the $50$ largest US market cap stocks. So we only build a portfolio by going long on the top $10$ stocks \cite{Baz2015}. In order to compare our methods with the performance of their stocks universe, we build a benchmark that consists of all 50 stocks weighted by their market cap. All portfolios are built including transaction costs.

We evaluate portfolio performance by calculating the Information ratio (IR), the ratio between excess return (portfolio returns minus benchmark returns) and tracking error (standard deviation of excess returns) \cite{IR}. We also calculate the downside information ratio, the ratio between excess return and the downside risk (variability of underperformance below the benchmark), that differentiates harmful volatility from total overall volatility. 

\section{Results}\label{sec:results}%\vspace{-0.3cm}
Tables~\ref{table:res_short_finbof} and~\ref{table:lstm_short} present the results obtained for each individual augmentation method and the combination of the most successful individual methods for the small 50 stock dataset using the LSTM and the TLo-NBoF networks. For comparison, we also show the results without augmentation. 
\begin{table}
\begin{minipage}[t]{0.48\textwidth}
\centering
\captionof{table}{Performance of the $k = 10$ long-only portfolios after transaction costs for the TLo-NBoF model and small dataset.}
\resizebox{\linewidth}{!}{
\begin{tabular}{rccccccc}
\toprule
{}               & Ann ret  & Ann vol & IR &  D. Risk & DIR & Acc   & F1 \\
\midrule
None            &   10.28 &   22.62 &  0.07  & 15.53 &   0.10  &  50.49$\pm$0.46 &     40.06$\pm$6.44 \\
Convolve        &   12.29 &   22.35 &  0.24  & 15.04 &   0.35  &   {\bf 50.62$\pm$0.6} &    {\bf 42.98$\pm$6.5} \\
Jitter          &     9.2 &   22.32 &  -0.02  &  15.32 & -0.02 &  50.43$\pm$0.59 &    42.5$\pm$6.71   \\
Magnify         &   {\bf 13.33} &  21.98 &  {\bf 0.31}   &  14.71 &   {\bf 0.47} &   50.55$\pm$0.5 &   40.35$\pm$6.35  \\
Pool            &   12.76 &    21.9 &  0.28  &   14.8 &   0.41  &   50.51$\pm$0.6 &    41.33$\pm$6.52 \\
Quantize        &   12.69 &   20.23 &  0.27  &  13.83 &   0.38  &  50.45$\pm$0.63 &   40.67$\pm$6.51  \\
Reverse         &    7.28 &   22.08 &  -0.18 &  15.03 &   -0.27  &  50.52$\pm$0.59 &   40.28$\pm$6.17 \\
Time warp       &   12.81 &   22.41 &  0.27 &  14.89 &   0.42  &  50.44$\pm$0.61 &   41.64$\pm$5.64  \\
Spawner         &  11.93  &   21.99 &  0.20  &  14.89 &  0.29 &  $50.46 \pm 0.53$ & $41.63 \pm 6.94 $ \\
Mag-Pool        &    9.24 &   22.58 &  -0.01 &  15.18 &   -0.02 &  50.52$\pm$0.44 &   40.2$\pm$6.68  \\
Mag-Quant       &   11.52 &   21.43 &  0.16 &  14.48 &    0.24  &  50.43$\pm$0.55 &   39.63$\pm$6.13  \\
Mag-TW          &    10.4 &    21.5 &  0.08 &  14.75 &    0.11  &  50.46$\pm$0.56 &   40.15$\pm$6.41  \\
Quant-Pool      &   11.52 & {\bf 20.15} &  0.15 &  {\bf 13.69} &   0.21  &  50.54$\pm$0.53 &   41.51$\pm$6.62  \\
Quant-TW        &   12.06 &    20.7 &  0.20 &  14.09 &   0.29   &  50.54$\pm$0.46 &   41.09$\pm$6.7  \\
\bottomrule
\end{tabular}
}
\label{table:res_short_finbof}
\end{minipage}
\end{table}

\begin{table}
\begin{minipage}[t]{0.48\textwidth}
\centering
\captionof{table}{Performance of the $k = 10$ long-only portfolios after transaction costs for the LSTM model and small dataset.}
\resizebox{\linewidth}{!}{
\begin{tabular}{rccccccc}
\toprule
{}               & Ann ret       & Ann vol & IR &  D. Risk & DIR & Acc   & F1 \\
\midrule
None       &   12.24 &   {\bf 24.05} &  0.22  &   {\bf 15.89} &   0.33  &   50.8$\pm$0.75 &  47.69$\pm$4.81 \\
Convolve   &   12.33 &   25.91 &  0.21  &   16.95 &   0.33   &  50.74$\pm$0.81 &  48.43$\pm$3.39 \\
Jitter     &   11.75 &   24.35 &  0.18  &   16.49 &   0.27 &  50.89$\pm$0.73 &  {\bf 48.86$\pm$2.87} \\
Magnify    &   14.16 &   25.44 &  0.32  &   16.58 &   0.51 &  {\bf 50.94$\pm$0.68} &   48.57$\pm$3.2 \\
Pool       &   11.81 &   26.15 &  0.18  &   17.15 &   0.27 &  50.86$\pm$0.77 &   48.5$\pm$3.49 \\
Quantize   &    12.80 &   24.41 &  0.26  &   16.46 &   0.38  &  50.93$\pm$0.79 &   48.6$\pm$2.82 \\
Reverse    &    6.12 &   24.12 &  -0.22  &   16.27 &   -0.33  &  50.76$\pm$0.78 &  45.96$\pm$4.74  \\
Time Warp &    {\bf 15.60} &   24.38 &  {\bf 0.43}  &   16.12 &   {\bf 0.67}   &  50.85$\pm$0.74 &  48.24$\pm$3.48  \\
Spawner   &  14.58      & 24.49 & 0.38  &  16.02 &  0.60 & $50.95 \pm 0.74$ & $48.36 \pm 3.66$ \\
Mag-Quant  &   13.70 &   25.82 &  0.29 &  16.74 &   0.47 &  50.92$\pm$0.67 &  48.43$\pm$3.29 \\
Mag-TW      &  14.00 &   25.66 &  0.31 &  16.63 &   0.49 &  50.88$\pm$0.67 &  48.45$\pm$3.06 \\
% TW-Quant   &  --      & -- &  --  &  -- &   -- &  -- & -- & -- \\
\\
% & & & & & & & & \\
\bottomrule
\end{tabular}
}
\label{table:lstm_short}
% \centering
\end{minipage}
\end{table}

We also show classification metrics (accuracy and F1) over the 25 data splits expressed by the mean and standard deviation. In both models, the classification accuracy improvement is very small with respect to no augmentation, and for F1 as well. But we see that both the IR and DIR improve using several augmentation methods. Magnify and time warp methods are consistently good performers, as well as spawner. For the TLo-NBoF, IR increases four times with respect of no method, and time warp on the LSTM model doubles the IR. We anticipated that the Reverse method would not be effective - and in both cases it decreases overall performance. Further, we note that he combination of two augmentation methods does not always improve performance. 

\begin{table}
\begin{minipage}[t]{0.48\textwidth}
\centering
\captionof{table}{Performance of the $k = 10$ long-short portfolios after transaction costs for the LSTM model and large dataset.}
\resizebox{\linewidth}{!}{
\begin{tabular}{rccccccc}
\toprule
{}        & Ann ret       & Ann vol & IR &  D. Risk & IDR & Acc   & F1  \\
\midrule
None      & $34.64$       & 28.43 &  1.22  &  18.78 &   1.84  & $51.0\pm1.0$       & $48.5\pm2.1$        \\
Convolve  & $32.60$       & 25.99 &  1.25  &  17.49 &   1.86  & $51.1\pm0.9$       & $49.2\pm2.0$       \\
Jitter    & $34.35$       & 25.3 &  1.36  &  16.69 &   2.06   & $51.0\pm1.0$       & $\mbf{50.0\pm1.0}$  \\
Magnify   & $\mbf{46.56}$ & 29.41 &  1.58  &  19.56 &   2.38  & $\mbf{51.2\pm0.9}$ & $48.8\pm2.7$        \\
Pool      & $36.18$       & 26.16 &  1.38  &  17.15 &   2.11  & $51.1\pm0.9$       & $49.4\pm2.2$       \\
Quantize  & $29.42$       & {\bf 25.48} &  1.15  &  {\bf 16.62} &   1.77  & $51.0\pm1.0$   & $48.8\pm2.1$ \\
Reverse   & $33.03$       & 26.34 &  1.25  &   16.9 &   1.95 & $51.0\pm1.0$       & $47.3\pm4.1$       \\
Time warp & $47.01$       & 29.26 &  1.61  &  19.17 &   2.45  & $51.2\pm0.9$ & $49.8\pm1.6$       \\
Spawner   & $38.08$       & 27.85 &  1.37  &  18.05 &   2.11 &  $51.1\pm1.0$       & $49.1\pm2.2$ \\
Mag-Jit   & $30.03$       & 27.59 &  1.09  &  18.62 &   1.61   & $51.1\pm1.0$       & $49.4\pm2.0$         \\
Mag-TW    & $44.03$       & 27.41 &  1.61  &  17.66 &   2.49  & $51.1\pm1.0$       & $49.6\pm1.4$       \\
% Mag-Spawner   &  --      & -- &  --  &  -- &   -- &  -- & -- & -- \\
TW-Pool   &  44.98      & 26.21 &  {\bf 1.72}  &  16.82 &   {\bf 2.67} &  $51.1\pm 0.9$ & $49.3 \pm 2.0$ \\
% TW-Spawner   &  --      & -- &  --  &  -- &   -- &  -- & -- & -- \\
TW-Jitter   &  22.47      & 25.94 &  0.87  &  17.71 &   1.27 &  $51.1 \pm 1.0$ & $49.3 \pm 1.8$ \\
\bottomrule
\end{tabular}
}
\label{table:lstm_large}
\end{minipage}
\end{table}

\begin{figure}
    \centering
    \includegraphics[width=0.95\linewidth]{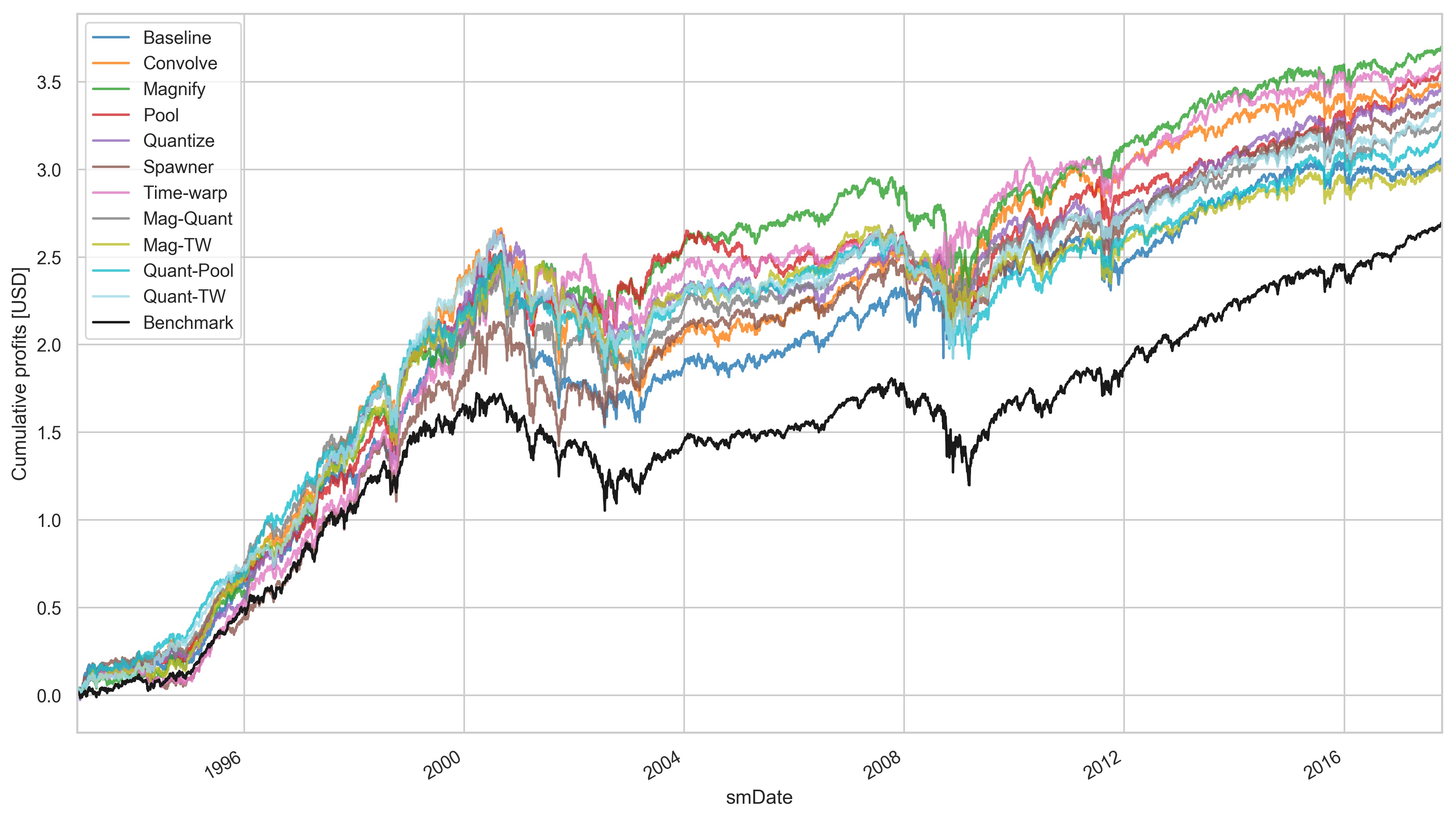}\\
    \includegraphics[width=0.95\linewidth]{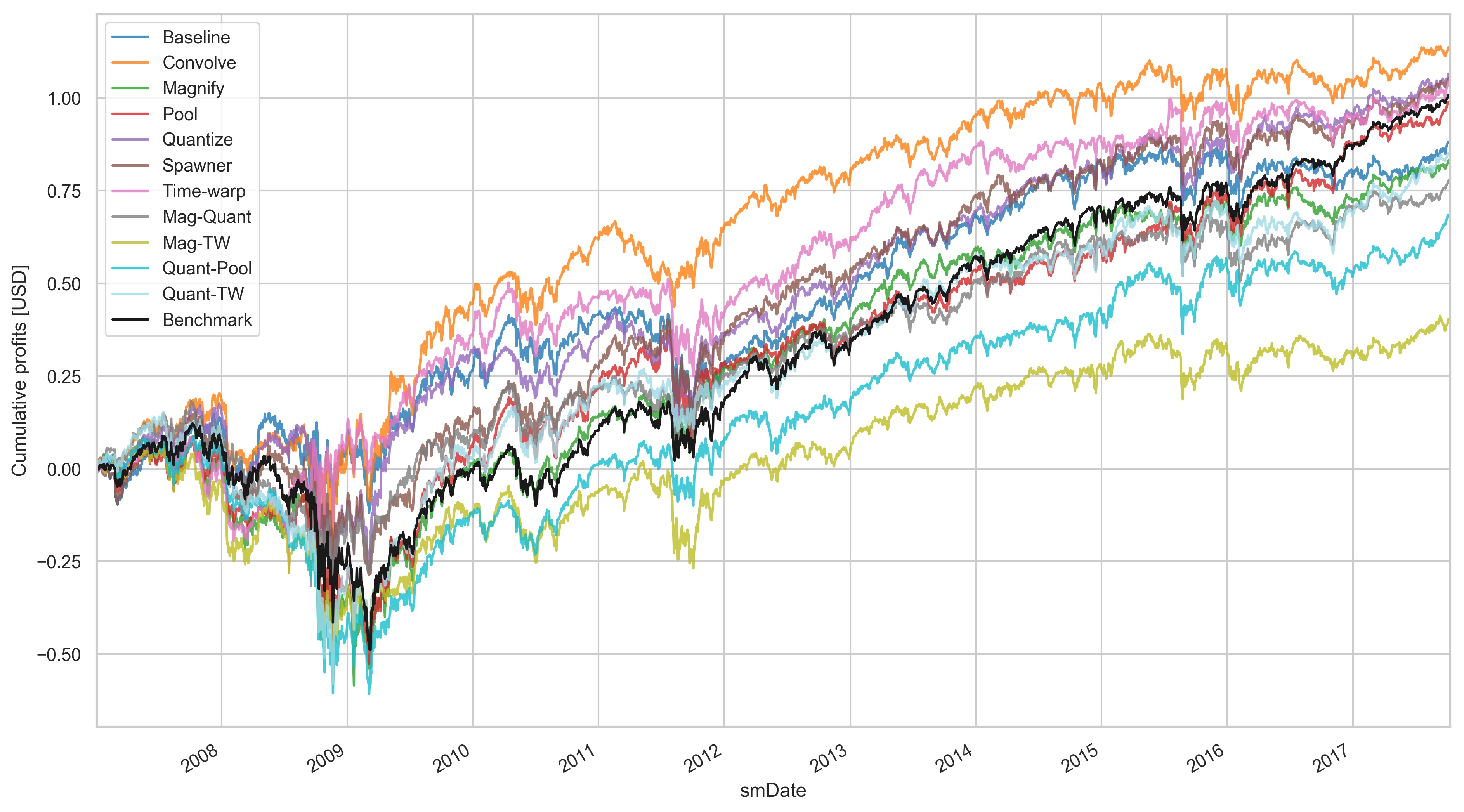}
    \caption{Performance of the TLo-NBoF models trained with and without augmentation and the benchmark (in black) measured as cumulative profits on 1USD average investment per day. Top corresponds to full testing history and bottom corresponds to the last 10 years.}
    \label{fig:res_short_finbof}
\end{figure}

\begin{figure}
    \centering
    \includegraphics[width=0.95\linewidth]{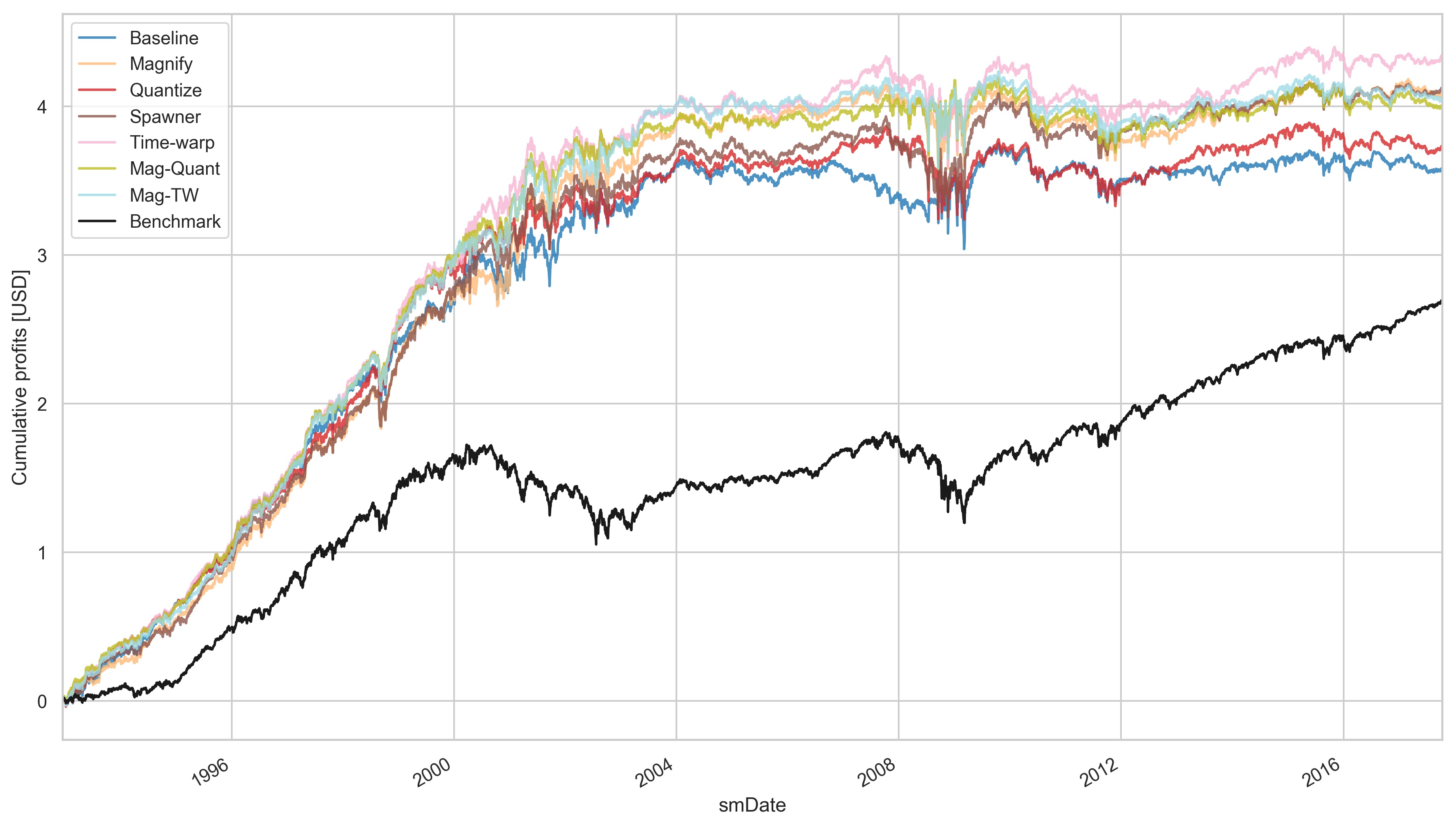}\\
    \includegraphics[width=0.95\linewidth]{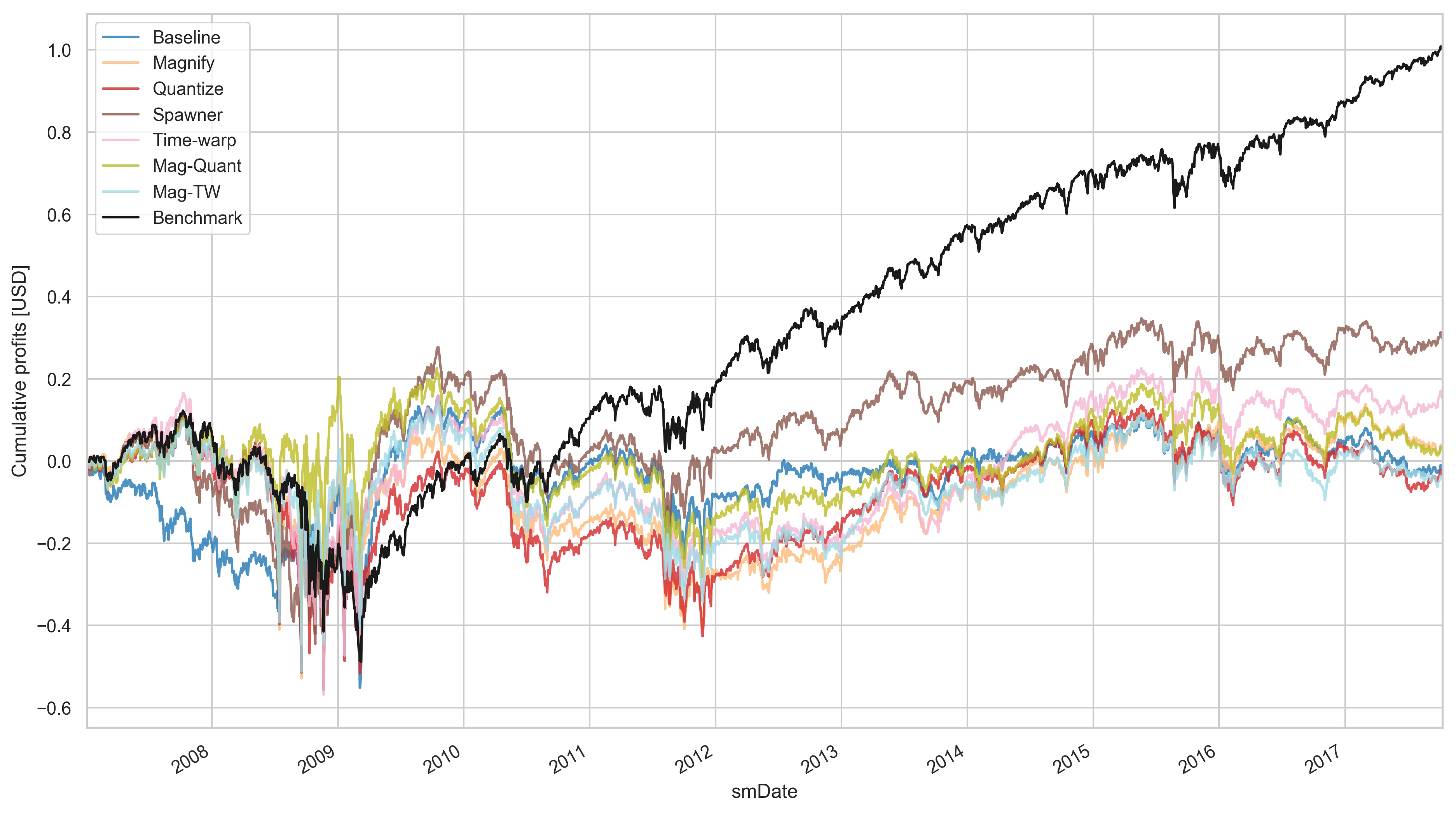}
    \caption{Performance of the LSTM models trained with and without augmentation and the benchmark (in black) measured as cumulative profits on 1USD average investment per day. Top corresponds to full testing history and bottom corresponds to the last 10 years.}
    \label{fig:res_short_lstm}
\end{figure}

Figures~\ref{fig:res_short_finbof} and~\ref{fig:res_short_lstm} show the cumulative profit over time (out of sample) of the models trained with different augmentation methods and the baseline (no augmentation). We focus on the most competitive techniques and for comparison, we add the benchmark calculated by the market weighted returns of the 50 constituent stocks. The top plots show the full history while the bottom plots show the last 10 years. Both models perform well over time, and cumulative profits of the models trained with augmentation are higher when compared to not using augmentation; however, only TLo-NBoF is competitive on the most recent testing period (2007-2017), along with several of the augmentation methods. The LSTM model fluctuates around zero and does not improve with regards to the benchmark. Krauss~\etal~\cite{Krauss2017} observes that the edge of the LSTM method seems to have been arbitraged away in the latter years. 
~

Table~\ref{table:lstm_large} presents the results obtained for each individual augmentation method and the combination of the most successful methods for the large S$\&$P500 dataset trained on the LSTM network. As the portfolios are long-short, they are market-neutral (therefore, the performance of the portfolio in independent of the performance of the market and no benchmark has to be subtracted). As with the small dataset, Magnify and Time warp show a strong performance in IR and DIR, as well as their combination. Jitter performs well in this dataset, but in the models trained on the small dataset, performance decreased so maybe in a larger dataset, the added noise helps with generalization, while in smaller data, diminishes the signal. The changes to the classification metrics are not significant. 

%\vspace{-0.2cm}
\section{Conclusions}\label{sec:conclusions}%\vspace{-0.2cm}

Data augmentation is a ubiquitous technique to improve generalization in supervised learning. In this work, we have studied the impact of various data augmentation methods for time series on the stock classification problem. We have shown that even with very noisy datasets such as stocks returns, it is beneficial to use data augmentation to improve generalization. Magnify, Time warp and Spawner consistently improve both the Information ratio and downside information ratio on all models and datasets. On the small datasets, augmentation achieves up to four-times (TLo-NBoF) and two-times (LSTM) performance improvement on IR compared to no augmentation. On a larger dataset, as espected, improvement is not that sharp, but still it achieves an increment in IR of up to $40\%$.

We tested the TLo-NBoF network that has not previously been used on low-freq stock data, and this network shows consistent positive returns over the last ten years of data, therefore, unlike the LSTM architecture, the profit has not been leveraged away.  

\bibliographystyle{IEEEbib}
\bibliography{biblio}

\begin{thebibliography}{10}

\bibitem{Iwana2020b}
Brian~Kenji Iwana and Seiichi Uchida,
\newblock ``Time series data augmentation for neural networks by time warping
  with a discriminative teacher,''
\newblock in {\em 2020 25th International Conference on Pattern Recognition
  (ICPR)}, 2020.

\bibitem{Teng2020}
Xiao Teng, Tuo Wang, Xiang Zhang, Long Lan, and Zhigang Luo,
\newblock ``Enhancing stock price trend prediction via a time-sensitive data
  augmentation method,''
\newblock {\em Complexity}, 2020.

\bibitem{Shorten2019}
Connor Shorten and Taghi~M. Khoshgoftaar,
\newblock ``A survey on image data augmentation for deep learning,''
\newblock {\em Journal of Big Data}, vol. 6, no. 1, pp. 60, 2019.

\bibitem{Wen2020}
Q.~Wen, Liang Sun, Xiaomin Song, J.~Gao, X.~Wang, and Huan Xu,
\newblock ``Time series data augmentation for deep learning: A survey,''
\newblock {\em ArXiv}, 2020.

\bibitem{Iwana2020}
Brian~Kenji Iwana and Seiichi Uchida,
\newblock ``An empirical survey of data augmentation for time series
  classification with neural networks,''
\newblock {\em arXiv preprint arXiv:2007.15951}, 2020.

\bibitem{guennec2016}
Arthur~Le Guennec, Simon Malinowski, and Romain Tavenard,
\newblock ``Data augmentation for time series classification using
  convolutional neural networks,''
\newblock in {\em ECML/PKDD Workshop on Advanced Analytics and Learning on
  Temporal Data}, 2016.

\bibitem{Fischer2018}
Thomas Fischer and Christopher Krauss,
\newblock ``{Deep learning with long short-term memory networks for financial
  market predictions},''
\newblock {\em European Journal of Operational Research}, vol. 270, no. 2, pp.
  654--669, 2018.

\bibitem{Cui2015}
X.~{Cui}, V.~{Goel}, and B.~{Kingsbury},
\newblock ``Data augmentation for deep neural network acoustic modeling,''
\newblock {\em IEEE/ACM Transactions on Audio, Speech, and Language
  Processing}, vol. 23, no. 9, pp. 1469--1477, 2015.

\bibitem{Goodfellow2016}
Ian Goodfellow, Yoshua Bengio, and Aaron Courville,
\newblock {\em Deep Learning},
\newblock MIT Press, 2016.

\bibitem{AlexK2012}
Alex Krizhevsky, Ilya Sutskever, and Geoffrey~E Hinton,
\newblock ``Imagenet classification with deep convolutional neural networks,''
\newblock in {\em Advances in Neural Information Processing Systems 25},
  F.~Pereira, C.~J.~C. Burges, L.~Bottou, and K.~Q. Weinberger, Eds., pp.
  1097--1105. 2012.

\bibitem{Simonyan2015}
Karen Simonyan and Andrew Zisserman,
\newblock ``Very deep convolutional networks for large-scale image
  recognition,''
\newblock in {\em International Conference on Learning Representations}, 2015.

\bibitem{Um2017}
Terry~T. Um, Franz M.~J. Pfister, Daniel Pichler, Satoshi Endo, Muriel Lang,
  Sandra Hirche, Urban Fietzek, and Dana Kuli\'{c},
\newblock ``Data augmentation of wearable sensor data for parkinson’s disease
  monitoring using convolutional neural networks,''
\newblock in {\em Proceedings of the 19th ACM International Conference on
  Multimodal Interaction}, 2017, ICMI '17, p. 216–220.

\bibitem{Baek2018}
Yujin Baek and Ha~Young Kim,
\newblock ``Modaugnet: A new forecasting framework for stock market index value
  with an overfitting prevention lstm module and a prediction lstm module,''
\newblock {\em Expert Systems with Applications}, vol. 113, pp. 457 -- 480,
  2018.

\bibitem{quantize2000}
Peter Tino, Christian Schittenkopf, and Georg Dorffner,
\newblock ``Temporal pattern recognition in noisy non-stationary time series
  based on quantization into symbolic streams. lessons learned from financial
  volatility trading.,''
\newblock Report Series SFB "Adaptive Information Systems and Modelling in
  Economics and Management Science"~46, SFB Adaptive Information Systems and
  Modelling in Economics and Management Science, WU Vienna University of
  Economics and Business, Vienna, 2000.

\bibitem{spawner}
Krzysztof Kamycki, Tomasz Kapuscinski, and Mariusz Oszust,
\newblock ``Data augmentation with suboptimal warping for time-series
  classification,''
\newblock {\em Sensors (Basel, Switzerland)}, vol. 20, no. 1, pp. 98, 12 2019.

\bibitem{arundo}
Arundo,
\newblock ``Tsaug,'' \url{https://tsaug.readthedocs.io/en/stable/index.html},
  2020.

\bibitem{Krauss2017}
Christopher Krauss, Xuan~Anh Do, and Nicolas Huck,
\newblock ``Deep neural networks, gradient-boosted trees, random forests:
  Statistical arbitrage on the $s\&p$ 500,''
\newblock {\em European Journal of Operational Research}, vol. 259, no. 2, pp.
  689 -- 702, 2017.

\bibitem{passalis2020}
Nikolaos Passalis, Anastasios Tefas, Juho Kanniainen, Moncef Gabbouj, and
  Alexandros Iosifidis,
\newblock ``Temporal logistic neural bag-of-features for financial time series
  forecasting leveraging limit order book data,''
\newblock {\em Pattern Recognition Letters}, 2020.

\bibitem{Baz2015}
Jamil Baz, N.~Granger, Campbell~R. Harvey, Nicolas~Le Roux, and Sandy Rattray,
\newblock ``Dissecting investment strategies in the cross section and time
  series,''
\newblock {\em Econometric Modeling: Derivatives eJournal}, 2015.

\bibitem{IR}
C.R. Bacon,
\newblock {\em Practical Risk-Adjusted Performance Measurement},
\newblock The Wiley Finance Series. Wiley, 2012.

\end{thebibliography}

\end{document}